\documentclass{aastex}
\usepackage{spr-astr-addons}
\usepackage{url}

\RequirePackage{color}

\begin{document}

\title{Synchrotron emission from the depths of pulsar magnetospheres}


\author{Z. Osmanov\altaffilmark{1}}
\affil{School of Physics, Free University of Tbilisi, 0183, Tbilisi,
Georgia email: z.osmanov@freeuni.edu.ge}
\author{Z. Yoshida\altaffilmark{2}}
\affil{Tokyo University, Tokyo, Japan}
\author{V.I. Berezhiani\altaffilmark{3}}
\affil{School of Physics, Free University of Tbilisi, 0183, Tbilisi,
Georgia}



\begin{abstract}
In this paper we study the generation of high energy emission from normal pulsars. For this purpose we consider the particles accelerated in the outer magnetosphere sliding along the closed magnetic field lines. It has been shown that  in due course of motion the initial small pitch angle increases and at a certain distance from the neutron star the synchrotron emission becomes significant. We found that a region covered from $150$ to $500$ stellar radii the emission pattern is characterised by energies in the interval $(0.1-10)$ MeV.
\end{abstract}



\section{Introduction}

One of the fundamental problems of modern pulsar astrophysics is the generation of electromagnetic waves in the high energy domain. It is believed that pulsar's radiation is produced mainly by means of the synchrotron mechanism \citep{pacini,shklovsky} and the inverse Compton scattering \citep{blandf}. On the other hand, one can straightforwardly show that typical timescales of synchrotron emission is so small compared to the escape timescale (which is of the order of rotation period of a pulsar) that almost from the very beginning of motion the particles go to the ground Landau state and after that slide along the magnetic field lines. This means that the overall dynamics of particles is described by a one-dimensional picture. 

It is observationally evident that some pulsars exhibit efficient radiation in the X-ray band \citep{x1,x2} and some of them usually belong to binary systems \citep{x3,x4}. It is clear that generation of radiation requires efficient acceleration of particles. In general it is thought that particles might accelerate from a nearby region of the polar caps (polar cap acceleration) \citep{polar1,polar2,polar3} as well as from the outer magnetosphere, by means of the so called outer gap mechanism \citep{outer1,outer2} or centrifugal acceleration \citep{or1,or2,or3}.

Already accelerated particles might generate electromagnetic radiation via several mechanisms: the synchrotron process, the inverse Compton scattering and the curvature radiation. In the magnetospheres of pulsars the synchrotron radiation power is so efficient that if the pitch angles are not very small the particles very soon transit to the ground Landau state and slide almost along the magnetic field lines, making dynamics almost one-dimensional. If the particles follow the so-called open field lines, the magnetic field decreases and the synchrotron process is strongly suppressed. Unlike the mentioned scenario, the particles which stay on closed field lines might turn back to the star's surface, potentially leading to a very interesting result. Due to the conservation of the adiabatic invariant, $I = 3cp_{\perp}^2/2eB$, where $c$ is the speed of light, $p_{\perp}$ is the transversal momentum of electron, $e$ is its charge and $B$ is the magnetic field induction, the value of $p_{\perp}$ \citep{landau} must inevitably increase if the particle slides along the field line and penetrates the region with strong magnetic field. On the other hand, the synchrotron emission is strongly dependent on the transversal component of momentum and as a result the corresponding emission mechanism will be revived inside the magnetosphere. 

If the particles are in frozen-in condition, they co-rotate with the field lines and consequently the force responsible for co-rotation is significant. On the other hand, the force responsible for radiation will be important if it exceeds the co-rotation force.

The paper is organised in the following way: in Sec. 2, we develop the model, and in Sec. 3 
we summarise our results for typical parameters of pulsars.

\section[]{Main approach}

By assuming that the particles are in frozen-in condition, they follow the magnetic field lines and thus dynamics of particles is described by a bead on the wire approximation. In the framework of the paper we consider the dipolar field lines when the vector of angular velocity of rotation, $\vec{\omega}$, and the dipolar magnetic momentum are perpendicular (see Fig. 1). Then, if one combines the prescribed dipolar channels, $\phi = \varphi (r)+\omega t$, with the Minkowskian metric defined in the polar coordinates
\begin{equation}
\label{metr1} 
ds^2 = -c^2dt^2+d\ell^2+\ell^2d\phi^2+dz^2,
\end{equation}
after some boring but straightforward algebra one obtains the metric on the co-rotating trajectories
\begin{equation}
\label{metr2} 
ds^2 = g_{00}(dx^0)^2+2g_{01}dx^0dx^1+g_{11}(dx^1)^2,
\end{equation}

\begin{equation}
\
g_{\alpha\beta}=
\begin{bmatrix}
    -1+\frac{\omega^2r^2}{c^2}m(r) & \frac{\omega r^2\varphi'(r)\sin\delta}{c} \\
    \\
   \frac{\omega r^2\varphi'(r)\sin\delta}{c}  & g^2(r)+k^2(r)+r^2h^2(r)m(r)
\end{bmatrix}
\
\end{equation}
where
$$\beta=\{0,1\}
$$

$$m(r) =1-\frac{r}{L}\cos^2\delta$$

$$g(r) = \frac{1-\frac{3r}{2L}\cos^2\delta}{\sqrt{1-\frac{r}{L}\cos^2\delta}}$$

$$k(r) = \frac{3}{2}\sqrt{\frac{r}{L}}\cos\delta$$

$$h(r) = \frac{\varphi'(r)\sin\delta}{1-\frac{r}{L}\cos^2\delta}$$

$\varphi (r)$ describes a shape of the dipolar magnetic field line in the proper plane ($r = L\sin^2\theta$), $L$ is the corresponding lengthscale, $\delta$ is the inclination angle of a particular field line's plane with resect to $\vec{\omega}$ and $x^0 = ct$, $x^1 = r$. 

As we have already mentioned we consider acceleration of particles close to the light cylinder (LC) surface (a hypothetical area where the linear velocity of rotation equals the speed of light). On the other hand, different planes of field lines, corresponding to different values of $\delta$ intersect the LC with different distances from the pulsar. It is clear that higher Lorentz factors will correspond to stronger magnetic fields, which is the case with $\delta = \pi/2$. For this reason we consider this particular case and make the relevant calculations.

By setting $\delta = \pi/2$ the polar radial coordinate, $\ell$, coincides with the proper radial coordinate of a particular field line, $r$, and consequently the aforementioned metric tensor reduces to
\begin{equation}
\
g_{\alpha\beta}=
\begin{bmatrix}
    -1+\frac{\omega^2r^2}{c^2} & \frac{\omega r^2\varphi'(r)}{c} \\
    \\
   \frac{\omega r^2\varphi'(r)}{c}  & 1+r^2(\varphi'(r))^2
\end{bmatrix}
\
\end{equation}

For this case the Lagrangian
\begin{equation}
\label{lag} 
L = -mc^2\left(-g_{00}-2g_{01}\frac{\upsilon}{c}-g_{11}\frac{\upsilon^2}{c^2}\right)^{1/2}
\end{equation}
does not depend on time explicitly, therefore, the Hamiltonian
\begin{equation}
\label{ham} 
\mathcal{H} = \upsilon\frac{\partial L}{\partial\upsilon}-L = \gamma mc^2\left(-g_{00}-g_{01}\frac{\upsilon}{c}\right) = E
\end{equation}
is a constant of motion and consequently the Lorentz factor behaves as \citep{or2}
\begin{equation}
\label{gamma} 
\gamma = \gamma_0\frac{1-\omega r_0^2\Omega_0/c^2}{1-\omega r^2\Omega/c^2},
\end{equation}
where $\upsilon = dr/dt$ is the radial velocity, $\Omega = \omega+\varphi'(r)\upsilon$ is the effective angular velocity of rotation and $\gamma_0$, $r_0$ and $\Omega_0 $ are the initial values of the Lorentz factor, the radial coordinate and effective angular velocity of rotation respectively.
From Eq. (\ref{ham}) one can straightforwardly show that the radial velocity behaves with $r$ as follows \citep{arsen}
$$v = c \frac{\sqrt{g_{11}+E^2}}{g_{12}^2+E^2g_{22}}\times$$
\begin{equation}
\times\left(-g_{12}\sqrt{g_{11}+E^2}\pm E\sqrt{g_{12}^2-g_{11}g_{22}} \right).
\end{equation}
Components of the effective force acting on the particle responsible for co-rotation is expressed as \citep{grg}
\begin{equation}
\label{Fr} 
F_r = \frac{d}{dt}\left(\gamma m\upsilon\right)-\gamma m\Omega^2r,
\end{equation}
\begin{equation}
\label{Ff}
F_{\varphi} = \frac{d}{dt}\left(\gamma m\Omega r\right)+\gamma m\Omega\upsilon,
\end{equation}
where $\gamma =\left(-g_{00}-2g_{01}\frac{\upsilon}{c}-g_{11}\frac{\upsilon^2}{c^2}\right)^{-1/2}$ is the relativistic factor of the particle and the time dependence is governed by the value of the radial acceleration \citep{grg}
\begin{equation}
\label{dv}
\frac{d\upsilon}{dt} = \frac{r\omega\Omega-\gamma^2r\upsilon\left(\varphi'+\omega\upsilon/c^2\right)\left(\Omega+r\varphi''\upsilon\right)}{\gamma^2\left(1-\omega^2r^2/c^2+r^2(\varphi')^2\right)}.
\end{equation}

Apart from the aforementioned force the charged particles also undergo the force providing the conservation of the adiabatic invariant
\begin{equation}
\label{Gpp} 
G_{\perp} = -\frac{c}{\rho}p_{\perp},
\end{equation}
\begin{equation}
\label{Gpr} 
G_{\parallel} = \gamma mc^2 \;\frac{p_{\perp}^2}{p_{\parallel}^2},
\end{equation}
where $p_{\perp}$ and $p_{\parallel}$ are respectively the transversal and longitudinal components of momentum and $\rho$ represents the curvature radius of the field lines. This force contributes in a very important process of maintenance of synchrotron emission despite the efficient energy losses. In particular, by means of the so called cyclotron instability the quasi linear diffusion leads to the following kinetic equation \citep{QLD}
\begin{equation}
\label{kin} 
\frac{\partial f}{\partial t}+\frac{1}{p_{\perp}}\frac{\partial}{\partial p_{\perp}}\left(p_{\perp}G_{\perp}f\right) = 
\frac{1}{p_{\perp}}\frac{\partial}{\partial p_{\perp}}\left(p_{\perp} D_{\perp,\perp}\frac{\partial f}{\partial p_{\perp}} \right),
\end{equation}
where 
\begin{equation}
\label{D} 
D_{\perp,\perp} \approx \frac{e^2mc^2n\gamma}{16\omega_B},
\end{equation}
is the corresponding diffusion coefficient, $n$ is the relativistic particles' number density and $\omega_B$ is the cyclotron frequency.

By considering the quasi stationary case $\partial/\partial t = 0$, one can straightforwardly obtain
\begin{equation}
\label{f} 
f(p_{\perp}) = C \exp\left(\int\frac{G_{\perp}}{D_{\perp,\perp}}dp_{\perp}\right) = Ce^{-\left(\frac{p_{\perp}}{p_{\perp,0}}\right)^2},
\end{equation}
where 
\begin{equation}
\label{pp0} 
p_{\perp,0} = \sqrt{\frac{2\rho D_{\perp,\perp}}{c}}.
\end{equation}
After averaging the value of $p_{\perp}$ one derives an expression for the residual pitch angle, $\psi_0 = p_{\perp,av}/(\gamma mc)$,
\begin{equation}
\label{pitch} 
\psi_0 = \frac{1}{\sqrt{\pi}}\frac{p_{\perp,0}}{\gamma mc}.
\end{equation}
Therefore, if the particles are accelerated close to the LC, the perpendicular momentum rapidly almost completely vanishes and a residual pitch angle defined by Eq. (\ref{pitch}) plays a role of the initial pitch angle. Usually these values are very small and consequently the synchrotron mechanism is strongly suppressed.

According to the standard approach, the particles accelerated on the open field lines, leave the magnetosphere and contribute in emission processes generated on the LC scales and beyond. On the other hand, there is a certain fraction of relativistic particles moving along the closed magnetic field lines. They sooner or later will  reach the inner areas of pulsar's magnetosphere and as a result, by means of the adiabatic invariant, the corresponding transversal component (pitch angle) will inevitably increase turning the synchrotron process back into the game. This in turn means that the synchrotron radiation reaction force having the following components \citep{landau}
\begin{equation}
\label{Hpp} 
H_{\perp} =-\alpha\;\frac{p_{\perp}}{p_{\parallel}}\left(1+\frac{p_{\perp}^2}{m^2c^2}\right),
\end{equation}
\begin{equation}
\label{Hpr} 
H_{\parallel} =-\alpha\;\frac{p_{\perp}^2}{m^2c^2}
\end{equation}
with $\alpha = 2e^2\omega_B^2/(3c^2)$ will become significant.

The major difference from the standard emission mechanisms is that we consider the trapped particles, accelerating in the outer gap region, but which are going back towards the neutron star's surface emitting from higher and higher magnetic field regions. The analysis of the aforementioned forces might reveal an area in the pulsar's magnetosphere, where the synchrotron mechanism will be efficient. This particular problem we consider in the following section.

\begin{figure}
  \centering {\includegraphics[width=8cm]{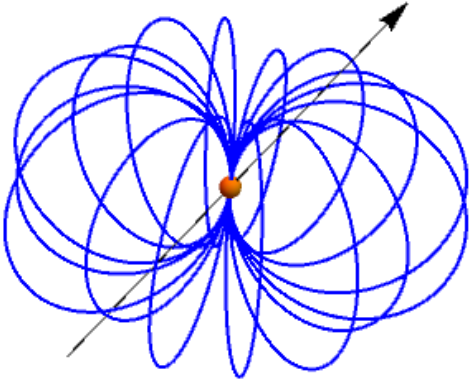}}
  \caption{Here we schematically show the pulsar's magnetosphere. The arrow indicates a direction of the vector of the angular velocity of rotation which is perpendicular to the magnetic dipole momentum.}\label{fig1}
\end{figure}

\begin{figure}
  \centering {\includegraphics[width=8cm]{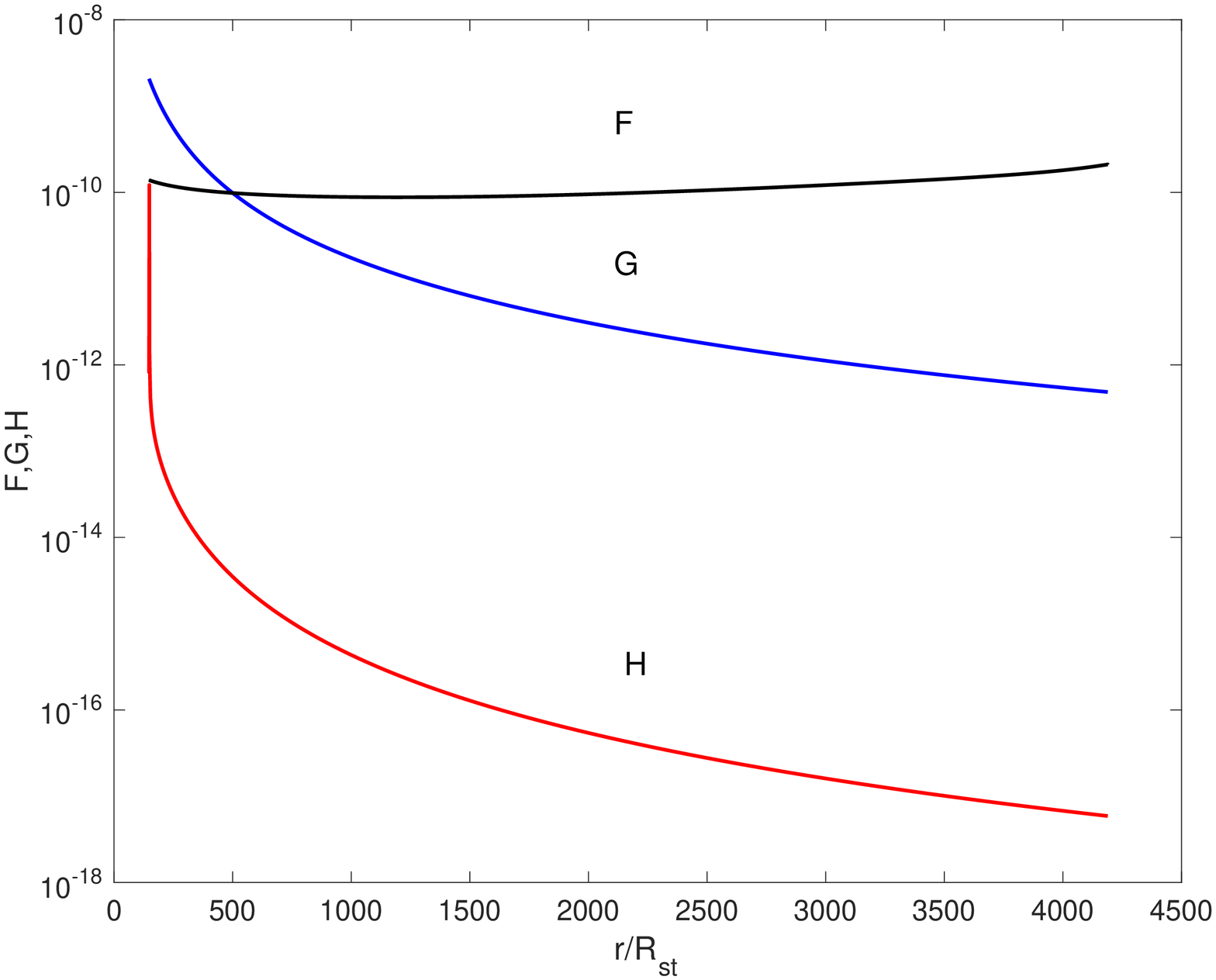}}
  \caption{Here we plot the forces, F (black), G (blue) and H (red) versus distance from the star's centre (normalised by the star's radius). The set of parameters is $P = 1$sec, $\dot{P} = 10^{-15}$ss$^{-1}$, $L = 0.9 R_{lc}$, $R_{st} = 10^6$cm, $\rho \simeq L$, $r_0\simeq L$ and $\gamma_0 = 10^6$.}\label{fig2}
\end{figure}

\begin{figure}
  \centering {\includegraphics[width=8cm]{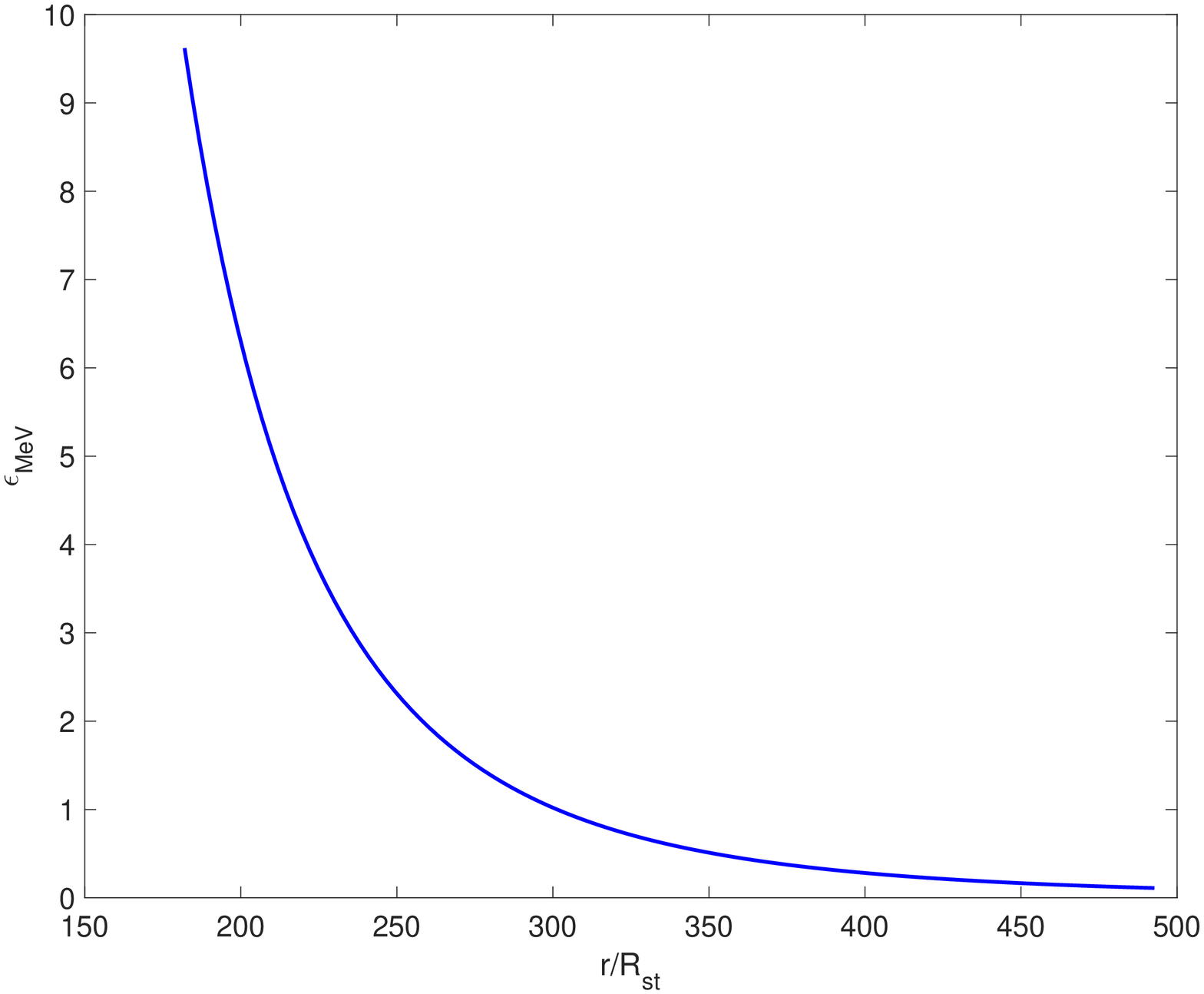}}
  \caption{In this plot we show the dependence of emission energy on distance (normalised by the neutron star's radius) is. The set of parameters is the same as in Fig. 1.}\label{fig3}
\end{figure}

\section{Discusion}
Here we examine normal pulsars with typical periods of rotation $P\simeq 1$sec, for which the magnetic field and the particle number density (the Goldreich-Julian density) are respectively given by
\begin{equation}
\label{B} 
B \simeq 10^{12}\times\left(\frac{P}{1s}\times\frac{\dot{P}}{10^{-15}erg/s}\right)^{1/2}\times\left(\frac{R_{st}}{r}\right)^3 \;G,
\end{equation}
and
\begin{equation}
\label{gj} 
n_{GJ} = \frac{\omega B}{2\pi ec}\frac{1}{1-\frac{\omega^2r^2}{c^2}}.
\end{equation}
where $\dot{P}\equiv dP/dt$, $R_{st} \simeq10km$ represents the neutron star's radius.

In this paper we consider the dipolar configuration of the field lines in polar coordinates $r = L\sin\varphi$, where $L$ is the scale parameter. In the framework of the approach we assume that particles are accelerated in the outer gap zone \citep{outer1,outer2}, where usually the Lorentz factors are of the order of $\gamma = 10^{6-7}$. In general it is believed that the mentioned area is very close to the LC, therefore the choice $L = 0.9 R_{lc}$ ($R_{lc}=c/\Omega$ is the light cylinder radius) might be considered as to be a reasonable parameter. 

On the other hand, we are interested in particles, following the field lines. The frozen-in condition works if the plasma energy density is small compared to the magnetic field energy density
\begin{equation}
\label{cond} 
\gamma mc^2n_{GJ} \leq\frac{B^2}{8\pi},
\end{equation}
leading to the following Lorentz factor
\begin{equation}
\label{gcor} 
\gamma \leq\frac{eB}{4\omega mc}\left(1-\frac{\omega^2r^2}{c^2}\right).
\end{equation}
By combining this expression with Eq. (\ref{B}), one can show that the highest value of the initial Lorentz factor still providing the frozen-in condition is of the order of $2\times 10^6$.

As a first example we consider $\gamma_0 = 10^6$, which gives the initial residual pitch angle of the order of $8\times 10^{-4}$. One can straightforwardly check that the force responsible for co-rotation exceeds all other forces, therefore, on the initial stage synchrotron emission is negligible. The electrons, sliding along the field lines, will move toward the star's surface, entering the higher magnetic field regions. As a result, by means of the adiabatic invariant the perpendicular momentum will increase
\begin{equation}
\label{pper} 
p_{\perp} = \frac{p_{\perp,0}}{\sqrt{\pi}}\left(\frac{r_0}{r}\right)^{3/2},
\end{equation}
leading to the increase of the value of $G$ (see Eqs. (\ref{Gpp},\ref{Gpr})). It is worth noting that if initially the plasma particles are frozen in strong magnetic fields this condition will be maintained in the whole course of motion. Indeed, when particles move along the field lines back to the neutron star's surface, by means of the centrifugal force it will decelerate, leading to the decrease of the Lorentz factor. On the other hand, the regions closer to the star have stronger magnetic fields. Therefore, if initially the particles follow the field lines, they will always stay on them.

In Fig. 2 we show the behavioour of the aforementioned forces, $F = \sqrt{F_{r}^2+F_{\varphi}^2}$ (black), $G = \sqrt{G_{\perp}^2+G_{\parallel}^2}$ (blue) and $H = \sqrt{H_{\perp}^2+H_{\parallel}^2}$ (red) versus the distance from the pulsar's centre. The distance is normalised by the neutron star's radius. The set of parameters is $P = 1$sec, $\dot{P} = 10^{-15}$ss$^{-1}$, $L = 0.9 R_{lc}$, $R_{st} = 10^6$cm, $\rho \simeq L$, $r_0\simeq L$ and $\gamma_0 = 10^6$. As it is clear from the plots, on the distance of the order of $500 R_{st}$ the force responsible for the adiabatic invariant, $G$, becomes comparable to $F$ and consequently the synchrotron emission becomes important from this area. On the distance of the order of $150 R_{st}$ the synchrotron emission becomes most efficient because as it is evident from the figure the radiation reaction force becomes of the same order of magnitude as the force responsible for co-rotation. The corresponding gamma ray energy of synchrotron emission is given by \citep{rybicki}
\begin{equation}
\label{eps} 
\epsilon = \frac{3\gamma^2 eB}{4h\pi mc}\sin\psi,
\end{equation}
where $h$ is the Plank'c constant and the pitch angle behaves with distance as (see Eq. (\ref{pper}))
\begin{equation}
\label{psi} 
\psi = \psi_0\left(\frac{r_0}{r}\right)^{3/2},
\end{equation}

In Fig 3 we show the behaviour of synchrotron frequency vs distance (normalised by the neutron star's radius). The set of parameters is the same as in Fig. 1. As it is clear from the plot, from $150 R_{st}$ to $500 R_{st}$ the synchrotron emission provides energies in the interval $(0.1-10)$MeV. The energy is a continuously decreasing function of distance, which is a natural result of Eqs. (\ref{eps},\ref{psi}). In particular, by increasing the distance the magnetic field induction as well as the pitch angle decrease leading to the following behaviour.

It is worth noting that a similar mechanism has been examined in the context of generation of infrared radiation in the binary system \citep{mst}. In particular, the authors have shown that the cyclotron instability by means of the quasi-linear difussion can explain emission of a neutron star's accretion flow in the frequency domain $\omega\simeq 10^{14}$Hz.

\section{Conclusion}
For studying the generation of high energy emission by means of the synchrotron mechanism we examine particles already accelerated in the outer magnetosphere. By means of the so-called quasi linear diffusion it has been shown that if the particles have Lorentz factors of the order of $10^6$, the pitch angle becomes very small $8\times 10^{-4}$.
 
In due course of motion, as the particles reach the stars' surface the pitch angle increases. We have shown that initially the force responsible for co-rotation is higher compared to other two forces. At the distance $500 R_{st}$  the value of $G$ becomes of the order of $F$, which means that the pitch angle becomes significant and the synchrotron emission comes into the game. 

By reaching the surface the pitch angles increase even more and at $150 R_{st}$ the radiation reaction force becomes comparable to $F$, implying that at the mentioned location the emission is extremely efficient. 

We have found that in the interval $150-500$ stellar radii the synchrotron emission might provide energies from $0.1$MeV to $\sim 10$MeV. 

Generally speaking, the considered process might be significant in millisecond pulsars as well as magnetars and therefore, a certain extension of the present work could be done.

\section*{Acknowledgments}
The research was supported by the Shota Rustaveli National Science Foundation grant (NFR17-587). ZO and VB also acknowledge hospitality of the Department of Advanced Energy at the University of Tokyo during their visit in 2019.
\bibliographystyle{spr-mp-nameyear-cnd}

\begin{thebibliography}{99}

\bibitem[Arsenadze \& Osmanov(2017)]{arsen} Arsenadze, G. \& Osmanov, Z., 2017, IJMPD, 26, 1750153
\bibitem[Blandford et al.(1990)]{blandf} Blandford R. D., Netzer H. \& Woltjer L., 1990, Active Galactic Nuclei.
Springer-Verlag, Berlin
\bibitem[Cheng \& Ruderman(1980)]{polar3} Cheng, A.F. \& Ruderman, M.A., 1980, ApJ, 235, 576
\bibitem[Coe et al.(2011)]{x3} Coe, M.J. et al., 2011, MNRAS, 414, 3281
\bibitem[Chkheidze et al.(2011)]{QLD} Chkheidze, N., Machabeli, G. \& Osmanov, Z., 2011, ApJ, 730, 12
\bibitem[Harding \& Muslimov(1998)]{polar2} Harding, A.K. \& Muslimov, A.G., 1998, ApJ, 508, 328
\bibitem[Hirotani(2015)]{outer1} Hirotani, K., 2015, ApJL, 798, 5
\bibitem[JaeSub et al.(2017)]{x1} JaeSub, H. et al., 2017, ApJ, 847, 26
\bibitem[Landau \& Lifshitz(1971)]{landau} Landau, L.D. \& Lifshitz, E.M. 1971, Classical Theory of Fields (London: Pergamon)
\bibitem[Machabeli et al.(1987)]{mst} Machabeli, G.Z., Sakhokia, D.M. \& Taktakishvili, A.L., 1987, PAZh, 13, 32
\bibitem[McGowan et al.(2007)]{x2} McGowan et al., 2007, MNRAS, 376, 759
\bibitem[Nobuyuki \& Shinpei(2003)]{polar1} Nobuyuki, S. \& Shinpei, S., 2003, ApJ, 584, 427
\bibitem[Osmanov \& Rieger(2019)]{or1} Osmanov, Z. \& Rieger, F.M., 2019, A\&A, 627, 5
\bibitem[Osmanov \& Rieger(2017)]{or2} Osmanov, Z. \& Rieger, F.M., 2017, MNRAS, 464, 1347
\bibitem[Osmanov \& Rieger(2009)]{or3} Osmanov, Z. \& Rieger, F.M., 2009, A\&A, 502, 15
\bibitem[Pacini(1971)]{pacini} Pacini F., 1971, ApJ, 163, 117
\bibitem[Rybicki \& Lightman(2007)]{rybicki} Rybicki, G.B. \& Lightman, A.P., Radiative Processes in Astrophysics, Wiley 2007
\bibitem[Rogava et al.(2003)]{grg} Rogava, A., Dalakishvili G. \& Osmannov Z., 2003, Gen.Rel.Grav., 35, 1133
\bibitem[Shklovsky(1970)]{shklovsky} Shklovsky I. S., 1970, ApJ, 159, L77
\bibitem[Sturm et al.(2011)]{x4} Sturm, R. et al., 2011, A\&A, 527, 7
\bibitem[Takata et al.(2008)]{outer2} Takata, J., Chang, H.K. \& Shibata, S., 2008, MNRAS, 386, 748




\end{thebibliography}

\end{document}